\documentstyle[prb,aps]{revtex}
\begin{document}
\draft
\title{First-Principles Elastic Constants for the hcp Transition Metals Fe, Co,
and Re at High Pressure}
\author{Gerd Steinle-Neumann and Lars Stixrude}
\address{Department of Geological Sciences, University of Michigan, Ann Arbor, MI 48109-1063}
\author{Ronald E. Cohen}
\address{Geophysical Laboratory and Center for High Pressure Research, Washington, DC 20015-1305}
\date{\today}
\maketitle
\begin{abstract}
The elastic constant tensors for the hcp phases of three transition metals
(Co, Re, and Fe) are computed as functions of pressure using the
Linearized Augmented Plane Wave method with both the local density and 
generalized gradient approximations.
Spin-polarized states are found to be stable for Co (ferromagnetic) and Fe
(antiferromagnetic at low pressure).
The elastic constants of Co and Re are compared to 
experimental measurements near ambient conditions 
and excellent agreement is found. 
Recent measurements of the lattice strain in high pressure experiments 
when interpreted in terms of elastic constants for Re and Fe 
are inconsistent with the calculated
moduli.
\end{abstract}
\pacs{62.20.Dc,62.50.+p,71.20.Be,71.15.Ap}

\section{Introduction}
\label{intr}

The effect of pressure on the propagation of elastic waves in materials is
essential for understanding interatomic interactions, mechanical stability
of solids, phase transition mechanisms, material strength, and the internal
structure of the Earth and other planets.
However, little is known of the elasticity of solids at high pressure.
The experimental study of the elasticity of materials under high pressure is 
challenging, as traditional methods have been applied only to moderate 
pressures.
Ultrasonic measurements are generally limited to a few GPa \cite{jack90},
while Brillouin spectroscopy has been applied up to 25 GPa \cite{shim95}.

We investigate the elasticity of three hexagonal transition metals at high 
pressure: iron, rhenium, and cobalt. High pressure properties of iron are of 
considerable geophysical
interest as the Earth's solid inner core is composed
primarily of this element. 
The elasticity of hcp iron is important for understanding the elastic anisotropy
of the inner core \cite{creager,tromp,stix95}, and its 
super-rotation \cite{song}.
Rhenium is the strongest metal known at high pressure \cite{jean91} 
and is widely used as a gasket material
in diamond anvil cell  experiments. 
We have chosen cobalt for this study because of its proximity to iron in the
periodic table and as an example of a ferromagnetic hcp metal.

All three of these metals have been studied experimentally 
under high pressure and their equations-of-state are well known.
Iron transforms from  
the bcc phase at ambient conditions to hcp near
13 GPa \cite{jeph86}; the equation-of-state of the hcp phase has 
been measured up to 300 GPa \cite{mao91}.
Recent advances in
diamond anvil cell techniques have made it possible to
evaluate the lattice strain in a polycrystal subjected to a non-hydrostatic 
stress field which can be associated with elastic constants. The elasticity 
of iron has been inferred by this method at high pressure 
(up to 210 GPa) \cite{singh98,mao98}.
The equation-of-state of cobalt has been measured up to 80 GPa 
\cite{fuji96} and its elastic constants were obtained at zero 
pressure using traditional ultrasonic methods\cite{schober}.
In the case of rhenium the equation-of-state is known 
to 215 GPa \cite{vohra87}, its elastic constants and their pressure derivatives
have been ultrasonically measured at low pressure\cite{mangh74}.
The same experimental method for evaluating lattice strains as in hcp iron
has been applied to rhenium in the pressure range 18-37 GPa \cite{duffy98}.

Iron has been studied widely with first-principles theoretical approaches
because of its geophysical importance and the well known 
failure of the local density approximation (LDA) to the exchange-correlation 
potential to predict the ferromagnetic bcc ground state \cite{wang85}.
This failure was a major impetus in the development of 
the generalized gradient approximation (GGA) \cite{PWI,PWII,PBE}.
The equation-of-state of hcp iron under LDA and GGA is well known to high
pressures\cite{bagno89,tera92,stix94,soed} and its elastic constants 
have been calculated by the full-potential linearized muffin-tin orbital
method (FP-LMTO)\cite{soed}, and a total energy tight-binding (TB) method 
\cite{stix95,cohen97}.
For hcp cobalt calculations 
have been performed with the LMTO method in the atomic sphere 
approximation for LDA \cite{min86} and the linearized combination of
atomic orbital method (LCAO) for GGA \cite{leung91}. 
There is no previous theoretical work on the elastic constants of
hcp cobalt.
For rhenium only one study has focused on the hcp phase at high pressure 
\cite{fast95}; using FP-LMTO with LDA the equation-of-state
and the elastic constants at zero pressure have been calculated. 

We organize the paper as follows. 
Section \ref{meth} elaborates the computational 
details of our first principles calculations and our approach to calculating 
the elastic constants, the elastic wave velocities, and the acoustic anisotropy.
It is followed by a section presenting 
our results on the magnetic state of the materials studied, their 
$c/a$ ratios, the equation-of-state, and the elastic constants 
as functions of pressure. 
We compare our results in terms of the elastic wave velocities to high pressure 
experiments and the Earth's inner core. 
In section \ref{dis} we analyze the elastic anisotropy resulting from 
our calculations,
recent experimental and theoretical results, and the 
predictions from a central nearest neighbor force model. 
Finally, we present our conclusions in section \ref{conc}.

\section{Method}
\label{meth}

\subsection*{LAPW Total Energy Calculations}

We investigate the energetics of hcp iron, cobalt, and rhenium using the
full-potential linearized-augmented plane-wave method (LAPW) \cite{singhbook}
with both LDA and GGA approximations to the exchange-correlation potential.
For LDA the form of Hedin and Lundquist \cite{HL} and von Barth and Hedin 
\cite{BH} are used for non-magnetic and spin-polarized calculations, 
respectively. For GGA we adopt the efficient formulation of Perdew, Burke, 
and Ernzhofer \cite{PBE}. 

Core states are treated
self-consistently using the full Dirac equation for the spherical part of the
potential, while valence states are treated in a semirelativistic approximation
neglecting spin-orbit coupling.
We investigate ferromagnetic alignment in spin-polarized calculations for
all metals and antiferromagnetism for iron.
For consistency of the results all parameters in the
calculations except for spin-polarization are kept fixed.
 
For the $3d$ metals $3s$, $3p$, $3d$, $4s$, and $4p$ states
are treated as valence electrons for all volumes. For rhenium we treat all electrons
up to $4f$ as core, $5d$ and $6s$ as valence states. For rhenium we also
have tested other
configurations, such as including the $4f$ as valence states,
which did not change our results significantly.
The muffin-tin radii
$R_{MT}$ are 2.0 Bohr for the $3d$ metals, and 2.3 Bohr for rhenium.
As spin-orbit coupling of the valence electrons is important for the band
structure and other properties
of heavy elements, we consider the influence of the spin-orbit
term on the equation-of-state for Re by including it in a 
variational step\cite{singhbook}.

We carry out total energy calculations over a wide range of volumes for all
three metals (0.7 $V_0$ - 1.2 $V_0$, with $V_0$ the zero pressure volume).
At each volume we determine the equilibrium ratio of the lattice constants 
$c/a$ by performing calculations
for several different values of this ratio.
The equation-of-state is obtained by describing the energy-volume curve with
a third order expansion in the Eulerian finite strain\cite{birch}.

We have performed convergence tests with respect to Brillouin zone sampling and 
the size of the basis set,
$R_{MT}K_{max}$, where $K_{max}$ is the largest reciprocal space wave-vector
in the basis set. 
Converged results are achieved with 
a 12x12x12 special k-point mesh \cite{monk}, yielding 114 k-points in
the irreducible wedge of the Brillouin zone for the hcp lattice, and up to
468 k-points for the monoclinic lattice used in elastic constants calculations.
The number of k-points in the full Brillouin zone is well above the
convergence criterion for elastic constant calculations 
established by Fast et al.\ \cite{fast95}.
The size of the basis set is given by $R_{MT}K_{max}=9.0$, yielding  158 to 311
basis functions, depending on volume.
Careful convergence tests show that with these parameters
relative energies 
are converged to better than $0.1$ mRy/atom,
 magnetic moments to better than $0.05$ $\mu_{B}$/atom,
and $c/a$ to within 0.025.

\subsection*{Elastic Constants}

We calculate the elastic constants as the second
derivatives of the internal energy with respect to the strain tensor 
($\varepsilon$).
We choose the applied strains to be isochoric (volume-conserving) 
which has several important
consequences: 
First, we assure the identity of our calculated
elastic constants with the stress-strain coefficients, which are
appropriate for the calculation of elastic wave velocities; this
identity
is non-trivial for finite applied pressure 
\cite{wall,BK65}.
Second, the total energy depends on the volume much more strongly 
than on strain; 
by choosing volume conserving strains we
obviate the separation of these two contributions to the total energy. Third,
the change in the basis set associated with the applied strain is minimized,
thereby minimizing computational uncertainties.

We obtain the elastic constants at the equilibrium relaxed structure 
at any volume $V$ by straining the lattice,
relaxing the symmetry allowed internal degrees of freedom, and evaluating the
total energy
changes due to the strain as a function of its magnitude $\delta$.

The bulk modulus $K$ is calculated by 
differentiating the equation-of-state. 
For hexagonal crystals $K$ is the combination of elastic constants
\begin{equation}\label{Kc}
K=\left[C_{33}\left(C_{11}+C_{12}\right)-2 C_{13}^2\right]/C_S,
\end{equation}
with 
\begin{equation}
C_S=C_{11}+C_{12}+2C_{33}-4C_{13}.
\end{equation}
The volume dependence of the optimized $c/a$ is related to the difference in the
linear compressibilities along the $a$- and $c$-axes ($k_a$ and $k_c$). 
The dimensionless quantity $R$ describes this as
\begin{equation}
R=K(k_a-k_c)=-\frac{d \ln(c/a)}{d \ln V}.
\end{equation}
In terms of the elastic constants,
\begin{equation}
R=\left(C_{33}-C_{11}-C_{12}+C_{13}\right)/C_S.
\end{equation}
We calculate $C_S$ by varying the $c/a$ ratio at a given volume,
according to the isochoric strain 
\begin{equation}
\varepsilon(\delta)=\left(\begin{array}{ccc}\quad \delta \quad& \quad 0 & \quad 0 \\ 
\quad 0 \quad & \quad \delta & \quad 0 \\
\quad 0 \quad & \quad 0 & \quad \left(1+\delta\right)^{-2}-1 \end{array}\right).
\end{equation}
The corresponding energy change is
\begin{equation}
E\left(\delta\right)=E\left(0\right)+C_SV\delta^2+O(\delta^3).
\end{equation}
In the expressions for $C_S$, $K$, and $R$, $C_{11}$ and $C_{12}$ occur
only as a sum. To separate
these constants we determine their difference, $C_{11}-C_{12}=2C_{66}$ 
by applying an orthorhombic strain, space group {\it Cmcm}.
For the strained lattice we use the two atom primitive unit cell, 
with the atoms in the 
Wyckoff position $4c$, coordinates ($y$,$-y$,1/4).
The strain is
\begin{equation}
\varepsilon(\delta)=\left(\begin{array}{ccc}
\quad \delta \quad & \quad 0 & \quad 0 \\  
\quad 0 \quad & \quad -\delta & \quad 0 \\ 
\quad 0 \quad & \quad 0 & \quad \delta^2/\left( 1 - \delta^2 \right)
\end{array}\right),
\end{equation}
leading to a change in total energy:
\begin{equation}
E\left(\delta\right)=E\left(0\right)+2C_{66}V\delta^2+O(\delta^4).
\end{equation}
In the unstrained lattice the atomic coordinate is $y=1/3$, but
varies under strain\cite{nastar95}. We relax our calculations with respect 
to this internal degree of freedom.

To determine $C_{44}$ we use a monoclinic strain, space group $C2/m$.
The atomic positions in the two atom primitive unit cell are $(1/6,5/6,1/4)$.
The strain applied
\begin{equation}
\varepsilon(\delta)=\left(\begin{array}{ccc}
0 & 0 & \delta \\
0 & \delta^2/\left( 1-\delta^2\right) & 0 \\ \delta & 0 & 0 
\end{array}\right)
\end{equation}
results in an energy change 
\begin{equation}
E\left( \delta \right)=E\left(0\right)+2C_{44}V\delta^2+O(\delta^4).
\end{equation}
The equilibrium positions of the atoms are unaffected by this strain and do
not need to be redetermined.\cite{nastar95}

While for $C_{66}$ and
$C_{44}$ the leading error term is of the order $\delta^4$, for $C_S$ it
is of third order in $\delta$. It is therefore crucial to include
positive and negative strains in the calculation for $C_S$.
The strain amplitudes applied are typically nine values of $\delta$ covering
$\pm$ 4\% for $C_S$; for $C_{66}$ and $C_{44}$, seven values of $\delta$ 
ranging to 6\% are applied.
The elastic constants are then given by the quadratic coefficient of 
polynomial fits to the total energy results; the order of the polynomial fit
is determined by a method outlined by Mehl \cite{mehl93}.

From the full elastic constant tensor we can determine the shear modulus $\mu$
according to the Voigt-Reuss-Hill scheme\cite{hill63a} 
and hence the isotropically
averaged aggregate velocities for compressional ($v_p$) and shear waves ($v_s$)
\begin{equation}
v_p=\sqrt{\left(K+\frac{4}{3}\mu\right)/\rho} \quad,\quad v_s=\sqrt{\mu/\rho},
\end{equation}
with $\rho$ the density.

More generally, the acoustic velocities are related to the elastic constants
by the Christoffel equation
\begin{equation}
\left( C_{ijkl}n_j n_k-M\delta_{il}\right)u_i=0,
\end{equation}
where $C_{ijkl}$ is the fourth rank tensor description of elastic constants, 
$\mathbf{n}$ is the propagation direction, 
$\mathbf{u}$ the polarization vector, 
$M=\rho v^2$ is the modulus of propagation
and $v$ the velocity.

The acoustic anisotropy can be described as
\begin{equation}
\Delta_i=\frac{M_i[{\mathbf n_x}]}{M_i[100]},
\end{equation}
where $\mathbf{n_x}$ is the extremal propagation direction other than $[100]$
and $i$ is the index for the three types of elastic waves (one longitudinal and
the two polarizations of the shear wave). 
Solving the Cristoffel equation for the hexagonal lattice one can calculate
the anisotropy of the compressional ($P$) wave as
\begin{equation}\label{1}
\Delta_P=\frac{C_{33}}{C_{11}}.
\end{equation}
For the shear waves the wave polarized perpendicular to the basal plane
($S1$) and the one polarized in the basal plane 
($S2$) have the anisotropies
\begin{equation}\label{2}
\Delta_{S1}=\frac{C_{11}+C_{33}-2C_{13}}{4C_{44}} \quad,\quad
\Delta_{S2}=\frac{C_{44}}{C_{66}}.
\end{equation}
While for $S2$- and $P$-waves the extremum occurs along the $c$-axis, 
for $S1$ it is at an angle of 45$^{\mbox{o}}$ from the $c$-axis in the 
$a$-$c$-plane.
We note that an additional extremum may occur for the compressional wave
propagation at intermediate directions depending on the values of the
elastic constants.

\section{results}
\label{res}

\subsection*{Magnetism}

We find a stable ferromagnetic state only in cobalt.
It is stabilized over a wide volume
range with the magnitude of the moment decreasing with pressure 
in agreement with previous theoretical
results on the
pressure dependence of magnetic moments \cite{stix94} in other 
transition metals. Only
at the smallest volume considered (50 Bohr$^3$, 180 GPa)
is the moment vanishingly small (Fig.\ \ref{mm}).
LDA and GGA yield consistent results and predict a
zero pressure magnetic moment of 1.55 $\mu_B$, in excellent
agreement with experiment (1.58 $\mu_B$ \cite{mey51}).

In the case of hcp iron, we also investigate two antiferromagnetic states.
The first consists of atomic layers of opposing spin
perpendicular to the $c$-axis (afmI). The other arranges the planes of opposite
spins normal to the $[100]$ direction in the hcp lattice; this can be described
by the 
orthorhombic representation of the hcp unit cell (space group $Pmma$)
with spin up in the $(1/4,0,1/3)$ and spin down in the $(1/4,1/2,5/6)$
position (afmII).
We find that both structures are more stable than the non spin-polarized 
state and that afmII is 
energetically favored over afmI. 
For both antiferromagnetic states the moment is strongly pressure dependent.
For afmI it vanishes at 
volumes larger than $V_0$ (Fig.\ \ref{mm}), in excellent agreement 
with results of Asada and Terakura \cite{tera92}.
The other structure, afmII, possesses a magnetic moment 
well into the stable pressure regime of hcp iron, up to $\sim$ 40 GPa.
(Fig.\ \ref{mm}). 
Because of frustration on the triangular lattice,
it is possible that more complex spin arrangements such as incommensurate
spin waves as for fcc iron \cite{uhl} or a spin glass are still more 
energetically favorable than afmII.

Diamond anvil cell in situ M\"ossbauer measurements of hcp iron\cite{taylor91}
have shown no evidence of magnetism in the hcp phase.
The low antiferromagnetic moment we calculate in the stable hcp regime and
the significant hysteresis of the bcc-hcp transition \cite{taylor91}
might explain that no magnetism in hcp iron has been detected in the high
pressure M\"ossbauer experiment.
In this context it may be relevant that indirect
evidence for magnetism exists at low pressure.  
Epitaxally grown iron-ruthenium superlattices have shown magnetism
occurring in hcp iron multilayers\cite{maurer91}. Its character, however,
is still controversial \cite{saint95,knab91}.

\subsection*{$c/a$ Ratios}

For all materials studied the $c/a$ ratio 
agrees with experimental data to within 2\% and
is essentially independent of the exchange correlation potential (GGA or LDA).
Equilibrium $c/a$ ratios for iron range from 
1.58 at zero pressure to 1.595 at 320 GPa. This is consistent with
experimental measurements\cite{jeph86,mao91} in the range of 15 to 300
GPa, which have shown considerable scatter. 
For cobalt the zero pressure $c/a$ ratio is calculated
as 1.615, increasing to 1.62 at a pressure of almost 200 GPa. The zero 
pressure $c/a$ is slightly lower than the experimental value 
of 1.623 \cite{mey51}. Diamond anvil cell experiments have found 
a higher value of $c/a$,
as much as the ideal value (1.633) \cite{fuji96}, 
this discrepancy might be due to the coexistence of hcp and
metastable fcc cobalt in the polycrystalline sample \cite{fuji96}. 
The $c/a$ ratio for rhenium (1.615) does not change over the
whole pressure range studied - and is in good agreement with 
experimental results (1.613)\cite{vohra87}.

\subsection*{Equation-of-State}

For the equation-of-state of rhenium, LDA shows 
better agreement with experimental data than does GGA
(Fig.\ \ref{EOS}, Table \ref{EOS_table}). 
GGA overestimates the zero pressure volume and softens
the bulk modulus, supporting a general pattern seen in prior density 
functional calculations 
using GGA for other $5d$ metals \cite{barb90,koer}.
Including spin-orbit coupling in the calculation has 
little effect on the
equation-of-state parameters, resulting in less than 1\% change in the zero 
pressure volume and 2\% in the bulk modulus.
For cobalt, as for other $3d$
metals GGA is superior to LDA and reproduces the experimental equation-of-state
to within 2\% in volume and 10\% in bulk modulus
(Fig.\ \ref{EOS}, Table \ref{EOS_table}). 

The discrepancy in the equation-of-state parameters of hcp iron
between non spin-polarized calculations and experiment is significantly 
larger than for the other
two metals studied here (Table \ref{EOS_table}) or other
transition metals \cite{barb90,koer}.
The zero pressure volume is underestimated by $\sim$ 9\%, and the 
zero pressure bulk modulus is too stiff by 75\% (Table \ref{EOS_table}).
Especially at low pressure the non-magnetic
equation-of-state deviates considerably from experimental values, 
while at high pressure the agreement is very good (Fig.\ \ref{EOS}).
The stabilization of antiferromagnetic states at low pressure 
can account for some of the discrepancy. 
For afmII magnetism persists to volumes smaller than $V_0$, 
resulting in a larger zero pressure volume, 
reducing the difference with experiment 
to 5\%, and lowering the bulk modulus considerably (Table \ref{EOS_table}). 
This is still larger than the difference in $V_0$ for cobalt and for
cubic iron phases ($<$ 3\%) \cite{stix94,soed}.
We attribute the remaining 
discrepancy between low pressure experimental data and the afmII 
equation-of-state (Fig.\ \ref{EOS})  
to the approximations in GGA and the possible 
stabilization of more complex spin arrangements than those considered here.

\subsection*{Elasticity}

The agreement of the calculated elastic constants for cobalt and rhenium 
with zero pressure experimental results\cite{schober,mangh74}
is excellent with a root mean square error
of better than 20 GPa for both metals and both exchange-correlation potentials
(Fig.\ \ref{cij}, Tables \ref{Co_cij} and \ref{Re_cij}). 
The initial pressure derivative of the elastic constants for rhenium 
is also well reproduced by the calculations (Fig.\ \ref{cij}).
LDA and GGA exchange-correlation potentials give almost equally good agreement,
the minor differences arising primarily from differences in the bulk modulus
(Tables \ref{EOS_table}, \ref{Co_cij}, and \ref{Re_cij}).

Our elastic constant calculations for rhenium and iron do not agree with the 
results of lattice strain experiments (Fig.\ \ref{cij}, Tables \ref{Re_cij} 
and \ref{Fe_cij}).
For rhenium the overall agreement between these experiments and our elastic
constants is better than for iron. $C_{11}$ and $C_{12}$ agree well over the
pressure range of the experiments, while the other longitudinal ($C_{33}$)
and off-diagonal constant ($C_{13}$) differ significantly (Fig.\ \ref{cij}, 
Table \ref{Re_cij})).  The shear elastic modulus ($C_{44}$) shows the largest 
discrepancy of all elastic constants (factor of 1.5).
For iron the results of the lattice strain experiments and our calculations 
are in reasonable agreement for the off-diagonal constants only.
The longitudinal moduli we obtain at 60 Bohr$^3$ and 50 Bohr$^3$($\sim$ 50
GPa and $\sim$ 200 GPa, respectively)
are larger by approximately 50\%. This is partly related to the overestimated 
bulk modulus in the calculations. The largest discrepancy, as in the case of 
rhenium, occurs in the shear elastic constants ($C_{44}$ and $C_{66}$). 

Aggregate properties such as the
bulk and shear modulus, and the compressional and shear wave velocity are
in somewhat better agreement between the theoretical results and the lattice 
strain experiment for both rhenium and iron (Figs.\ \ref{Re_v_p} and \ref{v_p}).
For rhenium, theory and experiment differ by less than 15\% in bulk and shear 
modulus (Fig.\ \ref{Re_v_p}).
For iron the discrepancy is considerable
at intermediate pressure but becomes smaller with increasing pressure, 
as already seen for the equation-of-state (Figs.\ \ref{EOS} and \ref{v_p}). At 
$\sim$ 200 GPa the difference in bulk modulus between GGA and experiment 
is less than 5\% and the elastic wave velocities differ by $\sim$ 10\%. 
The shear modulus differs by 25\% even at high pressure.

For iron the comparison with previous theoretical results gives a more 
coherent picture.
While the longitudinal elastic constants from our calculations are larger 
by 10-20\% compared to TB\cite{cohen97} 
and FP-LMTO results \cite{soed} (Table \ref{Fe_cij}), the elastic anisotropy
is similar: the pairs of longitudinal, shear, and off-diagonal
elastic moduli display similar values. 
For the TB study this is true over the whole pressure range considered, 
for the FP-LMTO calculations only at low pressure; the ratio of the 
off-diagonal constants ($C_{12}/C_{13}$)
is strongly pressure dependent in that study, varying from
0.9 at zero pressure to 0.6 at 400 GPa.

\section{Discussion}\label{dis}

We find that the elastic 
anisotropy (eqs.\ \ref{1} and \ref{2}) is similar for all 
three metals studied here. 
The magnitude of the anisotropy is 10$\pm$2\% for the longitudinal anisotropy
and $\Delta_{S1}$, and 30$\pm$3\% for $\Delta_{S2}$ and is nearly independent
of pressure (Fig.\ \ref{aniso}).
This is consistent with the experimentally observed behavior of other hcp
transition metals, all of which 
- except for the filled $d$-shell metals zinc and cadmium - show 
anisotropy of similar magnitude (Fig.\ \ref{aniso}).

These results can be understood by comparison to a hcp
crystal interacting with central nearest neighbor forces (CNNF) \cite{born}.
For this model the elastic anisotropy is independent of the interatomic
potential to lowest order in $P/C_{11}$, hence the anisotropy is dependent
on the symmetry of the crystal only. Born and Huang \cite{born} have shown
that from this CNNF model the elastic constants scale as 32:29:11:8:8
for $C_{33}$:$C_{11}$:$C_{12}$:$C_{13}$:$C_{44}$, yielding $\Delta_{P}=32/29$,
$\Delta_{S1}=8/9$, and $\Delta_{S2}=45/32$ (Fig.\ \ref{aniso}).

The experimentally determined elastic anisotropies of rhenium and
hcp iron at high pressure from lattice-strain measurements differ 
substantially from our theoretical predictions, previous theoretical
calculations, the behavior of all other hcp transition metals, and the
simple CNNF model (Fig.\ \ref{aniso}). 
The shear anisotropy in particular is very different in the high pressure
experiments as compared with all other relevant results. We suggest that this 
discrepancy may arise from assumptions made in the data analysis. In particular,
the assumption that the state of stress on all crystallographic planes is 
identical\cite{singh98}. This condition may not be satisfied in 
a material undergoing 
anisotropic deformation (e.\ g.\ dominated by basal slip), behavior that is
observed for many hcp transition metals.

Theory shows much better agreement with lattice-strain experiments in terms
of the isotropically averaged moduli. 
Even so, the agreement in the case of rhenium is much better than for iron.
In this context it is important to point out that part of the discrepancy in
the case of iron is due to the mutual inconsistency of the experimentally 
reported elastic constants and isotropic moduli. We have found that the
elastic constants reported in Ref.\ \onlinecite{mao98} do not yield the
values of $K$ and $\mu$ reported in the same paper. The reason for this 
discrepancy is unknown.

\section{Conclusions}
\label{conc}

The equations-of-state and the elastic constant tensor at zero pressure and
under compression for two ambient condition hcp transition metals, 
cobalt and rhenium, and for the high pressure phase of iron, hcp, are calculated
by means of the first principles LAPW method. We find a ferromagnetic ground
state for cobalt and an antiferromagnetic one for iron, with the 
antiferromagnetic moment vanishing at 60 Bohr$^3$. The equations-of-state
for the metals are in good agreement with experiment, as are the 
elastic constants and pressure derivatives of the elastic constants for 
cobalt and rhenium at ambient pressure. 

Elastic constants for iron under high pressure as inferred from lattice-strain 
experiments differ significantly from our theoretical results.
Similarly large discrepancies are also found between theory and high 
pressure static experiments on rhenium. 
The lattice-strain experiments 
also lead to large values of the shear anisotropy that differ from
that of all other open shell hcp transition metals.
Given the excellent agreement of the theoretical elastic constants for cobalt 
and rhenium with experiment at zero pressure,
we suggest that a re-examination of the lattice-strain experiments for rhenium 
and iron is warranted.

\section*{Acknowledgement}

We greatly appreciate helpful discussions with Tom Duffy, Rus Hemley, Dave Mao,
and Per S\"oderlind. The work was supported by the National Science Foundation
under grant EAR-9614790, and by the Academic Stategic Alliances Program of the  
Accelerated Strategic Computing Initiative (ASCI/ASAP) under subcontract no
B341492 of DOE contract W-7405-ENG-48 (REC). Computations were performed
on the SGI Origin 2000 at the Department of Geological Sciences at the 
University of Michigan and the Cray J90/16-4069 at the Geophysical Laboratory,
support by NSF grant EAR-9304624 and the Keck Foundation.


%
%
\begin{figure}
\caption{Magnetic moment per atom within the muffin-tin sphere for the 
two antiferromagnetic states of iron considered here and
the ferromagnetic moment for cobalt as a function of volume.}
\label{mm}
\end{figure}
\begin{figure}
\caption{Equations-of-state for the hcp metals considered. 
The upper panel compares the GGA non-magnetic (solid line) 
with the afmII structure (dotted line) for iron; 
Static experimental data is from
Ref.\ \protect\onlinecite{jeph86} (open circles) and Ref.\ 
\protect\onlinecite{mao91} (filled circles).
The lower two figures show the equations-of-state for ferromagnetic
cobalt and non-magnetic rhenium, GGA results are shown in solid, LDA in
dashed curves. The static experimental data for cobalt are from 
Ref.\ \protect\onlinecite{fuji96}, for rhenium static (open circles) and reduced 
shock wave data (filled circles) are from 
Ref.\ \protect\onlinecite{jean91} and Ref.\ \protect\onlinecite{marsh},
respectively. 
}
\label{EOS}
\end{figure}
\begin{figure}
\caption{
The elastic constants of hcp iron from our calculations are shown in the upper
figure. The lines are Eulerian finite strain fits to the theoretical results 
at 45, 50, and 60 Bohr$^3$: solid (GGA), dashed (LDA).
Lattice strain experiments
from Ref.\ \protect\onlinecite{singh98} and \protect\onlinecite{mao98} are
shown by the open symbols:
$C_{11}$ ({\large $\circ$}), $C_{33}$ ($\triangle$),  $C_{12}$
($\bigtriangledown$), $C_{13}$ ($\Diamond$), and $C_{44}$ ($\Box$).
In the middle panel elastic constants for hcp cobalt are shown 
as a function of volume. The curve is again a fit 
to the calculations at 65 , 70, and 75 Bohr$^3$. GGA is shown in
solid, LDA in dashed lines. At the zero pressure volume they are compared 
to ultrasonic experiments from Ref.\ \protect\onlinecite{schober} 
(filled symbols as above).
The lower figure shows the equivalent for rhenium with calculations 
at 85, 93, and 100 Bohr$^3$. 
The thick dotted lines indicate the initial pressure derivatives 
as determined from ultrasonic measurements
\protect\cite{mangh74}. For lattice strain experiments from Ref.\ 
\protect\onlinecite{duffy98} open symbols are used again.
}
\label{cij}
\end{figure}
\begin{figure}
\caption{Isotropic properties for hcp rhenium in comparison to 
experiments. The lower panel shows the bulk (K) and shear modulus ($\mu$)
of our calculations (GGA) in solid lines. The ultrasonic experiments at ambient 
condition from Ref.\ \protect\onlinecite{mangh74} are shown in filled circles
with the initial pressure dependence in thick dotted lines.
Lattice strain experiments from Ref.\ \protect\onlinecite{duffy98} are shown 
in open symbols.  The upper panel uses the same symbols as the lower one for 
the compressional ($v_p$) and shear wave velocity ($v_s$).
}
\label{Re_v_p}
\end{figure}
\begin{figure}
\caption{Bulk properties for hcp iron in comparison to experiments and the 
Earth's inner core. The lower panel shows the bulk (K) and shear modulus ($\mu$)
of our calculations (GGA) in solid lines. 
Diamond anvil cell experimental results are from Refs.\
\protect\onlinecite{singh98} ({\large $\bullet$}) and 
\protect\onlinecite{singh98} 
({\large $\circ$} and $\Box$, denoting two different approaches). 
Ultrasonic measurements in a multianvil experiment ($\Diamond$) are from 
Ref.\ \protect\onlinecite{singh98} as well.  The crosses display 
seismic observations of the inner core. The lower figure 
uses the same symbols as the upper one for the compressional ($v_p$) and
shear wave velocity ($v_s$). 
}
\label{v_p}
\end{figure}
\begin{figure}
\caption{As a measure of anisotropy the elastic constant ratios 
$C_{11}/C_{33}$, $(C_{11}+C_{33}-2C_{13})/4C_{44}$, and $C_{44}/C_{66}$,
which govern the compressional ($\Delta_{P}$) and shear
wave anisotropy ($\Delta_{S1}$ and $\Delta_{S2}$, respectively)
of the single crystal, are shown as a function of the number 
of d-electrons. The upper figure shows the shear elastic anisotropy 
$\Delta_{S1}$, the middle $\Delta_{S2}$, and the lower
the ratio of the longitudinal elastic constants $\Delta_{P}$. 
For all transition metals 
crystallizing in the hcp phase filled circles are used. 
The dashed lines show the CNNF model predictions. 
High pressure lattice strain results for iron from Ref.\
\protect\onlinecite{singh98} and for rhenium from Ref.\ 
\protect\onlinecite{duffy98} are displayed with gray squares. Our 
results are the open diamonds with the pressure dependence shown in solid
lines connected to the symbols.
}
\label{aniso}
\end{figure}
%
%
%
\begin{table}
\caption{Equation-of-state parameters from a third order finite Eulerian
strain expansion of the energy-volume relation for the hcp transition
metals. 
$V_0$, $K_0$, are the zero pressure volume and bulk modulus, respectively; 
$K_0^{'}$ the pressure derivative of the bulk modulus. For experimental 
values the bulk modulus is calculated from the elastic constants at
ambient pressure.
\label{EOS_table}} 
\begin{tabular}{llrrrr}
   &        & $E_0$ & $V_0$ & $K_0$ & $K_0^{'}$ \\
   &        & [Ry/atom]  & [Bohr$^3$] & [GPa]  \\
\tableline
Fe & exp \cite{mao91} & & 75.4 & 165 & 5.3 \\
   & LDA nm & -2541.1046 & 64.7 & 344 & 4.4 \\
   & GGA nm & -2545.6188 & 69.0 & 292 & 4.4 \\
   & GGA afmI & -2545.6195 & 70.5 & 210 & 5.5 \\
   & GGA afmII & -2545.6212 & 71.2 & 209 & 5.2 \\
   & LMTO GGA \cite{soed} & & 65.5 & 340 & \\
\tableline
Co & exp \cite{schober} & & 74.9 & 190 & 3.6(2) \\
   & LDA fm & -2782.1081 & 68.0 & 255 & 4.0 \\
   & GGA fm & -2786.7364 & 73.6 & 212 & 4.2 \\
   & LCAO GGA \cite{leung91} & & 76.2 & 214 & \\
   & LMTO LDA \cite{min86} & -2782.173 & 71.1 & 276 & \\
\tableline
Re & exp \cite{kittel,schober} & & 99.3 & 365 & \\
   & LDA nm & -33416.1921 & 98.2 & 382 & 3.9 \\
   & GGA nm & -33436.2502 & 103.0 & 344 & 3.9 \\
   & LMTO LDA \cite{fast95}& & 98.8 & 447 & 
\end{tabular}
\end{table}
\begin{table}
\caption{Elastic constants of hcp cobalt from theory (GGA, LDA) 
and experiment. $C_{66}=\frac{1}{2}\left( C_{11}-C_{12}\right)$ is added for
comparison with $C_{44}$.
\label{Co_cij}}
\begin{tabular}{ccccccc}
Volume & $C_{11}$ & $C_{33}$ & $C_{12}$ & $C_{13}$ & $C_{44}$ & $C_{66}$\\
 (Bohr$^3$) & (GPa) & (GPa) & (GPa) & (GPa) & (GPa) & (GPa) \\
\tableline
\multicolumn{7}{c}{{\it Ultrasonic Experiment (0 GPa)\cite{schober}}}\\
74.9 & 306 & 357 & 165 & 102 & 75 & 71 \\
\multicolumn{7}{c}{{\it GGA}}\\
75.0 & 325 & 365 & 165 & 105 &  90 & 80 \\
70.0 & 440 & 485 & 210 & 140 & 125 & 115 \\
65.0 & 580 & 640 & 290 & 195 & 160 & 145 \\
\multicolumn{7}{c}{{\it LDA}}\\
75.0 & 295 & 340 & 135 & 85 & 95 & 80\\
70.0 & 390 & 440 & 170 & 115 & 125 & 110\\
65.0 & 515 & 575 & 245 & 175 & 160 & 135\\
\end{tabular}
\end{table}
\begin{table}
\caption{Elastic constants of hcp rhenium from theory (present work: GGA, LDA) 
and experiment. $C_{66}=\frac{1}{2}\left( C_{11}-C_{12}\right)$ is added for
comparison with $C_{44}$.
\label{Re_cij}}
\begin{tabular}{ccccccc}
Volume & $C_{11}$ & $C_{33}$ & $C_{12}$ & $C_{13}$ & $C_{44}$ & $C_{66}$\\
 (Bohr$^3$) & (GPa) & (GPa) & (GPa) & (GPa) & (GPa) & (GPa) \\
\tableline
\multicolumn{7}{c}{{\it Ultrasonic Experiment (0 GPa) \cite{mangh74}}}\\
99.3 & 616 & 683 & 273 & 206 & 161 & 172\\
\multicolumn{7}{c}{{\it Lattice Strain Experiment (26.5 GPa)\cite{duffy98}}}\\
93.5 & 760(65) & 735(165) & 370(40) & 355(50) & 320(60) & 195(60) \\
\multicolumn{7}{c}{{\it GGA}}\\
100.0 & 640 & 695 & 280 & 220 & 170 & 180 \\
 93.0 & 815 & 900 & 385 & 300 & 205 & 215 \\
 85.0 & 1075 & 1200 & 555 & 435 & 265 & 260 \\
\multicolumn{7}{c}{{\it LDA}}\\
100.0 & 605 & 650 & 235 & 195 & 175 & 185 \\
93.0 & 780 & 855 & 350 & 280 & 200 & 215 \\
85.0 & 1040 & 1150 & 510 & 400 & 250 & 265 \\
\multicolumn{7}{c}{{\it FP-LMTO LDA \cite{fast95}}}\\
98.7 & 837 & 895 & 293 & 217 & 223 & 272
\end{tabular}
\end{table}
\begin{table}
\caption{Elastic constants for non-magnetic hcp Fe under compression 
(present work: GGA, LDA);
the pressure
range covered corresponds to approximately 50 GPa to 350 GPa, almost the
pressure in the Earth's center.
For comparison results of other studies at $\sim$ 60 Bohr$^3$ are included.
$C_{66}=\frac{1}{2}\left( C_{11}-C_{12}\right)$ is added for
comparison with $C_{44}$.
\label{Fe_cij}}
\begin{tabular}{ccccccc}
Volume & $C_{11}$ & $C_{33}$ & $C_{12}$ & $C_{13}$ & $C_{44}$ & $C_{66}$\\
 (Bohr$^3$) & (GPa) & (GPa) & (GPa) & (GPa) & (GPa) & (GPa) \\
\tableline
\multicolumn{7}{c}{{\it Lattice Strain Experiment (50 GPa)\cite{singh98}}}\\
60 & 640(55) & 650(85) & 300(55) & 255(40) & 420(25) & 170(55)\\
\multicolumn{7}{c}{{\it GGA}}\\
60 &  930 & 1010 & 320 & 295 & 260 & 305 \\
50 & 1675 & 1835 & 735 & 645 & 415 & 475 \\
45 & 2320 & 2545 & 1140 & 975 & 400 & 590 \\
\multicolumn{7}{c}{{\it LDA}}\\
60 &  860 &  950 & 280 & 260 & 235 & 290 \\
50 & 1560 & 1740 & 720 & 595 & 415 & 420 \\
45 & 2210 & 2435 & 1090 & 915 & 535 & 560 \\
\multicolumn{7}{c}{{\it Tight-Binding\cite{cohen97}}}\\
60 & 845 & 900 & 350 & 340 & 235 & 245 \\
\multicolumn{7}{c}{{\it FP-LMTO GGA\cite{soed}}}\\
60 & 870 & 810 & 255 & 320 & 235 & 310 \\
\end{tabular}
\end{table}
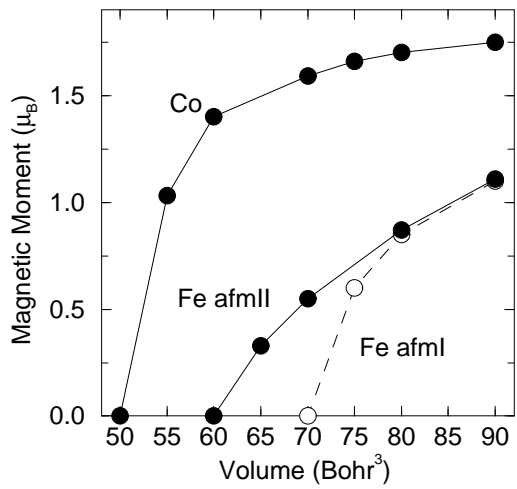
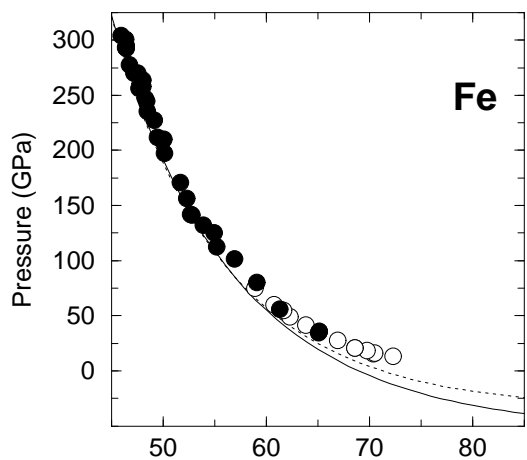
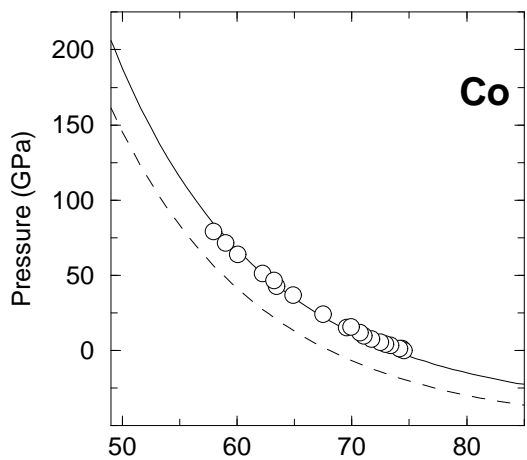
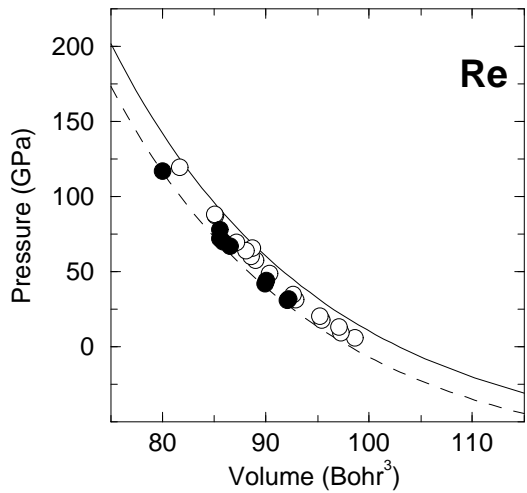
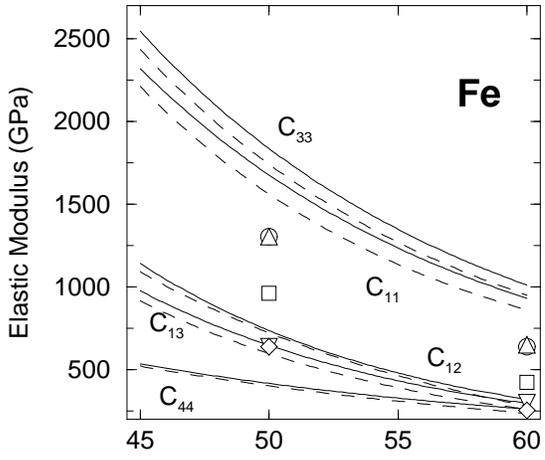
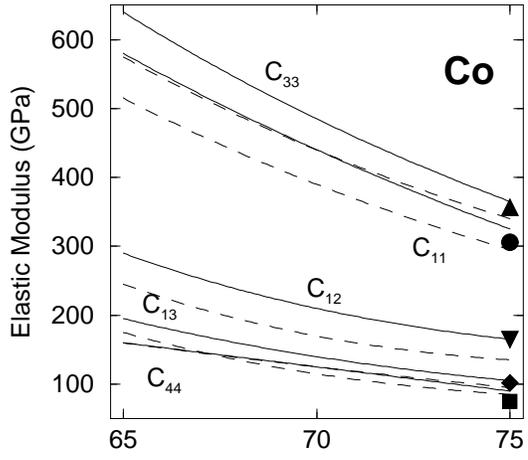
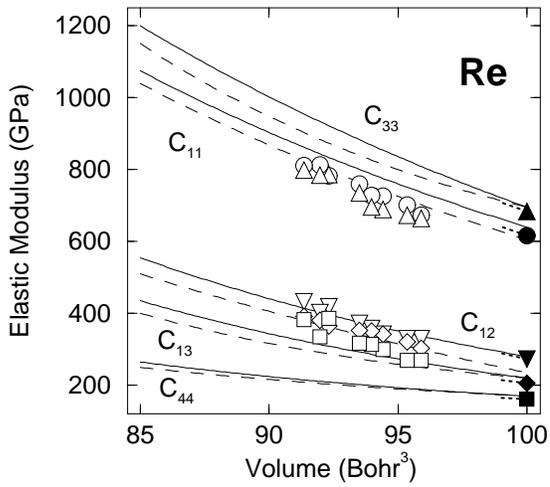
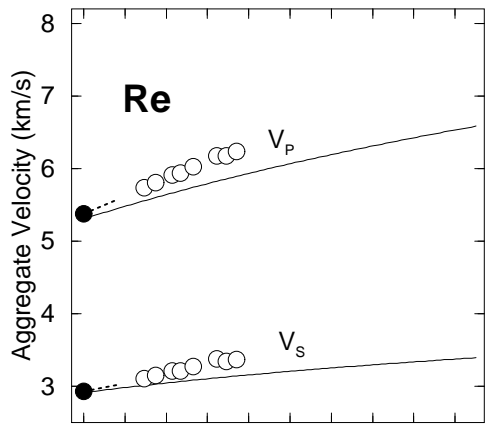
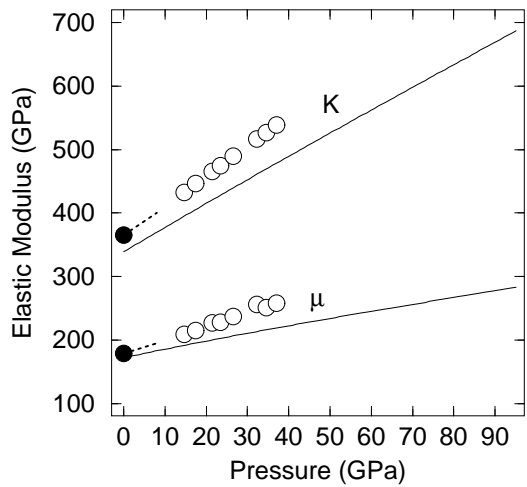
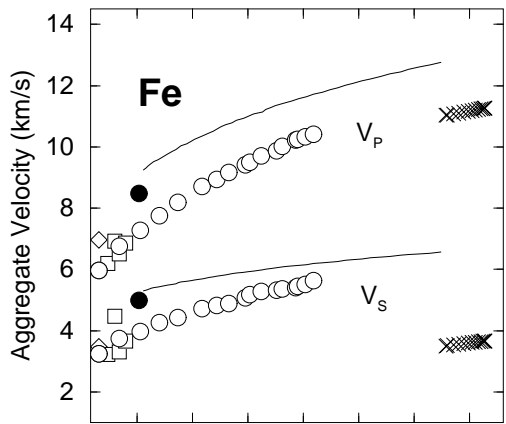
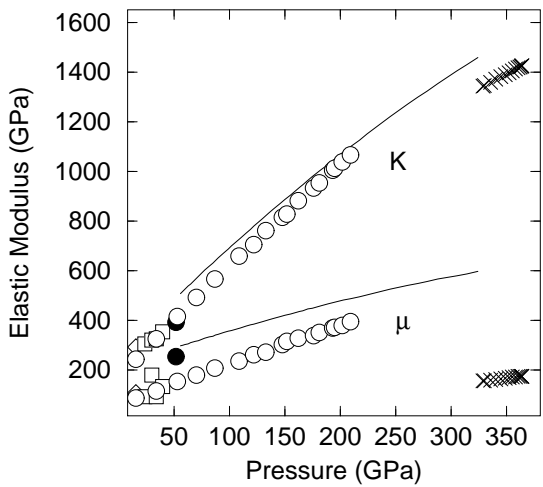
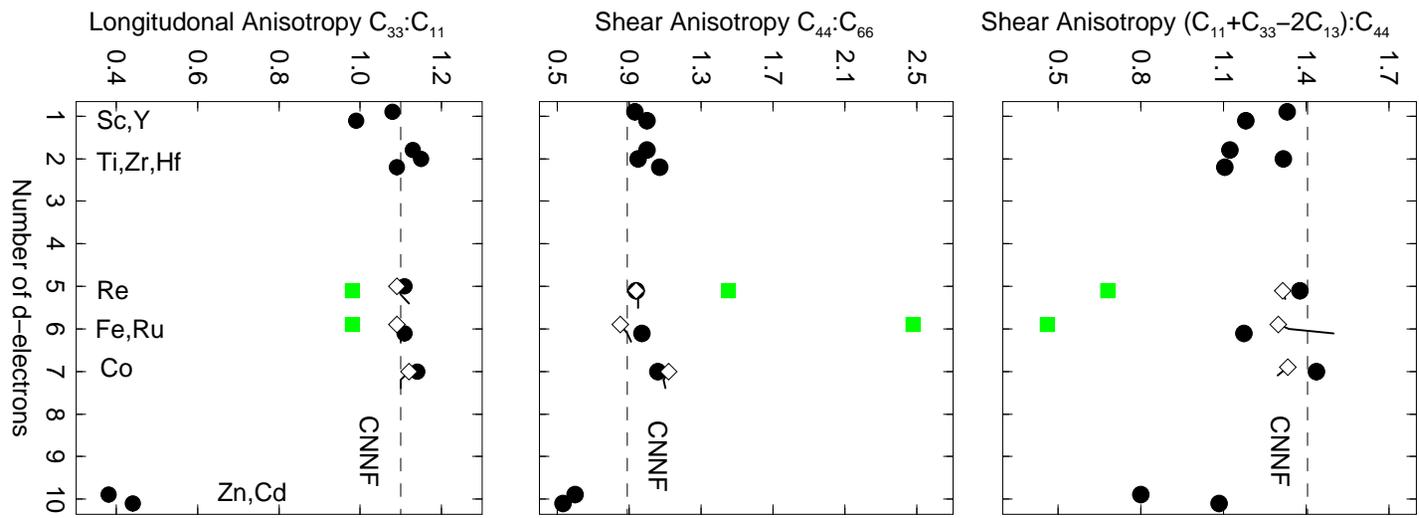
\end{document}